\documentclass[12pt]{article}
\usepackage{epsfig}
\begin{document}

\def\IThost{he117-90.jinr.ru}

\thispagestyle{empty}

\begin{center}
{\Large{\bfseries
% New software for the Nuclotron internal target
%control and data acquisition system
 New software of the control and data acquisition system
for the Nuclotron internal target station
}}\\[10mm]
{\Large{\bfseries A.Yu.~Isupov}
}\\[10mm]
{\itshape Veksler and Baldin Laboratory of High Energy Physics}\\[5mm]
{\itshape Joint Institute for Nuclear Research}
\end{center}

%\vspace*{2cm}

\newpage

\centerline{\bfseries Abstract}
%\label{inttarg.Abstract}

\vspace*{5mm}

\noindent Isupov~A.Yu. \\
%New software for the Nuclotron internal target
%control and data acquisition system
 New software of the control and data acquisition system
for the Nuclotron internal target station

\vspace*{5mm}

The control and data acquisition system for the internal target station (ITS)
of Nuclotron (LHEP, JINR) is implemented.
The new software is based on the
{\itshape ngdp} framework under the
Unix-like operating system FreeBSD to allow easy
network distribution of the online data collected from ITS,
% internal target and accompanying detectors,
as well as the internal target remote control.
%The revised CAMAC hardware contains now the JINR manufactured
%generic modules only.
%, while the too specific modules of the previous version are eliminated.

The investigation has been performed at the Veksler and Baldin Laboratory
of High Energy Physics, JINR.

\newpage

%\tableofcontents

%\newpage

\setcounter{page}{1}

\hfill\begin{minipage}{0.4\textwidth}
{\itshape Dedicated to the bright memory of V.A.Krasnov}
\end{minipage}

\section{Motivation and markup}
\label{inttarg.intro}

\vspace*{-2mm}

\hspace*{4mm} The current version of the Nuclotron internal target station
(ITS) is described in \cite{newIntTarg}.
% (see Fig.~\ref{inttarg.fig.ITS}).
The stepper motor and microstepper driver are still the same after the earliest
ITS version \cite{oldIntTarg}.

%\begin{figure}[htb]
%\epsfig{width=\textwidth
%,bbllx=0pt,bblly=0pt,bburx=576pt,bbury=430pt
%,file=.eps}
%\includegraphics[bb=0 0 576 430,width=\textwidth]{.eps}
%\caption{The general view of the current ITS version.}
%\label{inttarg.fig.ITS}
%\end{figure}

\begin{figure}[htb]
\epsfig{width=\textwidth
,bbllx=14pt,bblly=17pt,bburx=582pt,bbury=309pt
,file=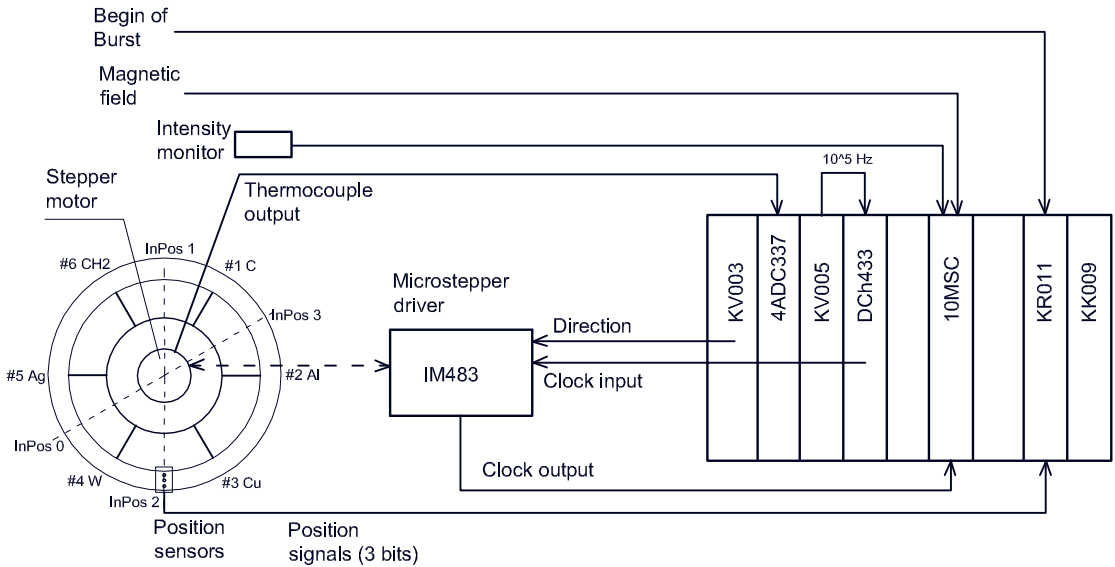}
\caption{Functional scheme of the internal target control.}
\label{inttarg.fig.func_scheme}
\end{figure}

During the 2010-11 years the
%The presented implementation of the
% Nuclotron internal target
ITS control system was reimplemented to achieve the following goals:\\
$\bullet$ replacement of outdated DOS software, which does not support
either the network, or the underlying computer hardware;\\
$\bullet$ replacement of the specific (almost unique now) CAMAC modules
by the more generic ones with higher availability and repairability;\\
$\bullet$ integration of
% keeping compatibility with
the already implemented {\bfseries\itshape targinfo(1)} server, which as a
workaround was used separately (under the SCAN DAQ \cite{AfanPTE08}) from the
internal target DOS software.

 The present paper is focused on the software of
the new ITS control and data acquisition (IntTarg CDAQ) system.
The CAMAC hardware being used now by the IntTarg CDAQ is shown in
Fig.~\ref{inttarg.fig.func_scheme}, where we can see JINR manufactured
generic modules only.

Through the present text the
% references to terms are highlighted as {\bfseries boldface text},
file and software package names are highlighted as
{\itshape italic text}, C and other languages constructions and reproduced
``as is'' literals --- as
\verb|typewriter text|. Reference to the manual page named ``qwerty'' in the
9-th section is printed as {\bfseries\itshape qwerty(9)}.
%,
%reference to the sections in this paper --
%as ``\ref{inttarg.sw.itGUI}''.
Note also verbal
constructions like ``{\bfseries\itshape accept(2)}ed'' and ``\verb|mkpeer|ing'',
which mean ``accepted by {\bfseries\itshape accept(2)}'' and
``peer making by \verb|mkpeer|''.
Subjects of substitution by actual values are
enclosed in the angle brackets: \verb|<cnts_mask>|, while some optional
parameters
%elements
are given in the square brackets: \verb|[-b<#>]|.

\section{IntTarg CDAQ software}
\label{inttarg.sw}

\subsection{{\itshape ngdp} based design}
\label{inttarg.sw.ngdp}

\hspace*{4mm}
As it was noted in \cite{IsupJINRC10-35}, the {\itshape ngdp}
framework could
be used to organize and manage the data streams originated, in particular,
from the CAMAC hardware. To reach some independence on the CAMAC crate
controller type, we use the current version of the {\itshape camac} package
%which predecessor is described in
\cite{GrOllat}. The {\itshape ngdp} using
allows us to eliminate intermediate data storage on slow media like hard disks
(HDD), as well as to gracefully distribute the data acquisition system
between more than one
networked computers if needed \cite{IsupJINRC10-34}. Of course, the
{\itshape ngdp} usage essentially reduces the
implementation efforts, as it is shown in
\ref{inttarg.sw.ngdp_mods}.

In the presented design of the user context utilities we adhere to the
{\itshape ngdp} and its predecessor {\itshape qdpb} \cite{IsupJINRC01-116}
%practice
convention when some command-line flags have
the same meanings for many utilities, as follows:

\begin{description}
\item[{\ttfamily -l}] Write logging information by
{\bfseries\itshape syslogd(8)}, facility \verb|LOG_LOCAL0| (may be changed
while compiling), levels \verb|LOG_ERR| and \verb|LOG_WARNING| instead
of the standard error output.
\item[{\ttfamily -v}] Produce verbose output instead of the short one by default.
\item[{\ttfamily -b<\#>}] Deal with the module, attached to the \verb|#|-th
branch, instead of the 0-th by default.
\item[{\ttfamily -p<pidfile>}] At startup write the own process identifier (PID)
in \verb|<pidfile>|. \verb|-p-| means to use a compiled-in default for
\verb|<pidfile>|.
\item[{\ttfamily -h}] Write the utility usage to the standard error
output and exit successfully.
\end{description}
So, in the specific utilities description we will not mention these flags.
Note also, that each utility exits 0 if it is successful
% on success
and nonzero --- if an error occurs.

\subsection{Ready modules used by IntTarg CDAQ}
\label{inttarg.sw.ngdp_mods}

\hspace*{4mm}
In the IntTarg CDAQ design we
% essentially
use the following
already implemented entities (introduced by the {\itshape ngdp} framework,
if not stated otherwise):\\
$\bullet$ The {\bfseries\itshape ng\_ksocket(4)} and {\bfseries\itshape ng\_socket(4)}
node types are standard in the {\bfseries\itshape netgraph(4)} package.\\
$\bullet$ The {\bfseries\itshape ng\_camacsrc(4)} node type (see \cite{IsupJINRC10-35})
allows us to inject data packets from a CAMAC interrupt handler (see
\ref{inttarg.sw.inttarg4}) into {\bfseries\itshape netgraph(4)} as data
messages.\\
% The {\bfseries\itshape ng\_camacsrc(4)}
%node type is introduced by the {\itshape ngdp} framework.\\
$\bullet$ The {\bfseries\itshape ng\_fifos(4)} node type (see \cite{IsupJINRC10-35})
implements the ``selfflow''\footnote{
Without internal bufferization and therefore request-free.
} queue with First Input First Output (FIFO) discipline
and is able to:\\
$\dag$ spawn \verb|listen()|ing {\bfseries\itshape ng\_ksocket(4)} at
startup;\\
$\dag$ spawn \verb|accept()|ing {\bfseries\itshape ng\_ksocket(4)}(s) at
each connection
request from the known host(s) / port(s) up to the configured maximum, and/or\\
$\dag$ accept hook connection from the local
{\bfseries\itshape ng\_socket(4)}(s);\\
$\dag$ emit each data packet obtained on the \verb|input| hook
as soon as possible (ASAP)
through all \verb|accept()|ing {\bfseries\itshape ng\_ksocket(4)}(s) and local
{\bfseries\itshape ng\_socket(4)}(s) currently connected;\\
$\dag$ close \verb|accept()|ing {\bfseries\itshape ng\_ksocket(4)} at
EOF notification obtaining or connection loss.\\
%The {\bfseries\itshape ng\_fifos(4)}
%node type is introduced by the {\itshape ngdp} framework.\\
$\bullet$ The {\bfseries\itshape ngget(1)} (see \cite{IsupJINRC10-35}) is
a utility
for the packet stream extraction from {\bfseries\itshape netgraph(4)} (usually
through the {\bfseries\itshape ng\_socket(4)} node type).\\
%and introduced by the {\itshape ngdp} framework.\\
$\bullet$ The {\bfseries\itshape writer(1)}
is a utility for
packet stream writing into regular files on HDD, and introduced by the
{\itshape qdpb} system \cite{IsupJINRC01-116}.
% (predecessor of the {\itshape ngdp}).

\subsection{Modules specific for IntTarg CDAQ}
\label{inttarg.sw.specific}

\subsubsection{CAMAC kernel module {\bfseries\itshape inttarg(4)}}
\label{inttarg.sw.inttarg4}

\hspace*{4mm}
The \verb|inttarg| module is intended to work with the
% internal target
ITS CAMAC hardware%
% (see \ref{inttarg.hw.camac})%
%, which controls the internal target station of the%
%Nuclotron (LHEP, JINR). The \verb|inttarg| module%
, complies with requirements
of the {\bfseries\itshape camacmod(9)} and {\bfseries\itshape ng\_ca\-mac\-src(4)},
so, it contains the CAMAC interrupt handler function. This handler
recognizes the following interrupt occurrences (events):\\
$\bullet$ begin of burst (BoB),\\
$\bullet$ target arrival to the initial position (InPos), and\\
$\bullet$ target departure from it.\\
%In whole
In total 8 packet types
%\verb|INTTARG_RUN_BEG|, \verb|INTTARG_RUN_END|, \verb|INTTARG_CYC_BEG|,
%\verb|INTTARG_CYC_END|, \verb|INTTARG_CYC_END1|, \verb|INTTARG_CYC_END2|,
%\verb|INTTARG_CYC_END3|, and \verb|INTTARG_INF_0|
\verb|INTTARG_CYC_{BEG,END,END[123]}|, \verb|INTTARG_INF_0|, and
\verb|INTTARG_RUN_{BEG,END}|
are produced. Note the
\verb|INTTARG_DAT_0| is not produced at all, because trigger events are
absent in the present design. All 4 \verb|END| packets have a variable length,
while all the others --- the fixed one.

At each BoB occurrence the \verb|inttarg| produces the \verb|INTTARG_CYC_BEG|
packet, which contains at least the time stamp (\verb|struct timeval|) of
the BoB.

If the target should be active during the current burst, the corresponding
(per-quantum) {\bfseries\itshape callout(9)} handler is established to be
executed once per our time quantum (compiled-in default is 10 msec,
usually it equals to 10 OS ticks). After each quantum this
handler increments the index in the trajectory description array already
supplied by {\bfseries\itshape itoper(8)}, reads the member
at this index, resets the effective
microstep frequency and direction according to this member value, collects
the experimental data from the ADC, and reestablishes itself.
After the final quantum the per-quantum handler produces the \verb|INTTARG_CYC_END2|
packet (ADC data) and wakes up the kernel thread
{\bfseries\itshape kthread(9)} to read the memory buffer of the 10MSC multiscaler
and produce the \verb|INTTARG_CYC_END1|
(magnetic field from the 9th 10MSC up/down input), \verb|INTTARG_CYC_END3|
(the 0..7th 10MSC regular inputs) and \verb|INTTARG_CYC_END| (target trajectory
from the 8th 10MSC up/down input) packets. Note that \verb|INTTARG_CYC_END| packet
always is the latest data packet of the current burst, while
\verb|INTTARG_CYC_END[123]| are
optional (depending on the software configuration).

The \verb|INTTARG_CYC_END| packet
% has variable length and
contains up to
$1+$\verb|arr_size|$+1$ of \verb|int16_t| values.
%The number of \verb|int16_t| values contained in the \verb|INTTARG_CYC_END|
%packet is up to $1+$\verb|arr_size|$+1$.
The \verb|arr_size| is a basic size of internal arrays in the \verb|inttarg|
and configured by
{\bfseries\itshape itconf(8)}, valid values are:
100..500, by default 500. The first (0th) \verb|int16_t| value contains the
\verb|union inttarg_cyc_end_qdt|
(see {\itshape einttarg.h}), which
has the \verb|quaval|, \verb|dtype| and \verb|tnum| fields. The
\verb|quaval| is a time quantum duration (10 msec), the \verb|dtype|
is a data type (valid values are
%\verb|INTTARG_CYC_END_INACTIVE|,
%\verb|INTTARG_CYC_END_MKSTEPS|, \verb|INTTARG_CYC_END_1_10MM|, and
%\verb|INTTARG_CYC_END_INVALID|,
\verb|INTTARG_CYC_END_{INACTIVE,MKSTEPS,1_10MM,INVALID}|,
see {\itshape einttarg.h}), and
the \verb|tnum| is a current target number (1..6). For our
case of the active target the \verb|dtype| is equal to
\verb|INTTARG_CYC_END_MKSTEPS| (or \verb|1_10MM|, if the
\verb|inttarg| is compiled without 10MSC support and the reported trajectory is
calculated instead of the really read-out one). Each other \verb|int16_t| value
is a signed
microstep's number of the stepper motor during the corresponding time quantum.
The positive values mean the movement from InPos, the negative ones ---
to InPos.

The \verb|INTTARG_CYC_END1| packet
% has variable length and
contains up to
$1+$\verb|arr_size|$+1$ of \verb|int16_t| values. The first (0th) is the
\verb|union inttarg_cyc_end1_qf| (see {\itshape einttarg.h}). Each other
\verb|int16_t| value is a signed magnetic field difference during the
corresponding time quantum. Naturally, the positive values correspond to the
field increasing, the negative ones --- decreasing.

The \verb|INTTARG_CYC_END2| packet
% has variable length and
contains up to
$1+$\verb|ADC_CHANS|$*($\verb|arr_size|$+1)$ of \verb|uint16_t| values.
The first (0th) equals to \verb|ADC_CHANS| constant ($=4$, means the number
of ADC channels, numbered from the 0th to the 3rd). After the first
\verb|uint16_t| the \verb|ADC_CHANS|
\verb|uint16_t| values represent the 1st time quantum, next \verb|ADC_CHANS|
\verb|uint16_t| values --- the 2nd time quantum, etc., and \verb|ADC_CHANS|
\verb|uint16_t| values for the \verb|arr_size|th final quantum. The 1st ADC
channel is modified to obtain the thermocouple
output to control the stepper motor temperature. The ADC value to temperature
conversion is as follows:
T ($^{\circ}$C)$ = -0.625 \times $\verb|ADC|$ + 628.75$.

The \verb|INTTARG_CYC_END3| packet
%has variable length and
contains up to
$1+$\verb|msc10nch|$*($\verb|arr_size|$+1)$ of \verb|uint32_t| values.
The first (0th)
%\verb|int32_t| value contains
is
the \verb|union inttarg_cyc_end3_cnt| (see {\itshape einttarg.h}), which
contains, in particular, the \verb|msc10nch| and \verb|msc10mask|
fields. The \verb|msc10nch| is a number (valid are 0..8) and the
\verb|msc10mask| is an 8-bit mask (valid are 0..\verb|0xff|) of the used 10MSC
regular inputs. The mask could be configured by {\bfseries\itshape itconf(8)}.
After the first \verb|uint32_t| the \verb|msc10nch| arrays with the same length
(up to \verb|arr_size|$+1$ members) contain the \verb|uint32_t| values of the
10MSC counts.
%the up to \verb|arr_size|$+1$ \verb|uint32_t|
%values are counts which represents
%the 1st used input, next
%the same number of
%% up to \verb|arr_size|$+1$
%\verb|uint32_t| values --- the 2nd used input, etc., and
%the same number of
%% up to \verb|arr_size|$+1$
%\verb|uint32_t| values for the \verb|msc10nch|-th used input.

If the target should be inactive during the current burst, the end-of-burst
(EoB) {\bfseries\itshape callout(9)} handler is established to be
executed after \verb|arr_size|
of 10 msec time quanta. The EoB handler
produces only the \verb|INTTARG_CYC_END| packet with an empty trajectory. This
means that the packet contains the \verb|union inttarg_cyc_end_qdt| only,
where the \verb|dtype| is equal to \verb|INTTARG_CYC_END_INACTIVE|.

The \verb|INTTARG_RUN_BEG| and \verb|INTTARG_RUN_END| are produced at
\verb|start| and \verb|stop| user requests (see {\bfseries\itshape itoper(8)})
and contain the
% corresponding
time stamps (\verb|struct timeval|).
% same fields as \verb|INTTARG_CYC_BEG| does.

The \verb|INTTARG_INF_0| packet is produced to indicate the operation in progress
or some error or warning condition, and it should be interpreted by the receiver
(for example, {\bfseries\itshape itGUI(1)} utility).
The \verb|INTTARG_INF_0| packet
contains: \verb|int32_t| value of the operation code (see
{\itshape toper\_op.h}), \verb|int16_t| value of the error or warning code
(see {\itshape inttarg\_err.h} and {\bfseries\itshape errno(2)}), and
\verb|int16_t| value of the attribute (for example, the current target number
at the \verb|WARN_CHTARG| warning).

The \verb|inttarg| module can be configured by the
{\bfseries\itshape itconf(8)} and
controlled by the {\bfseries\itshape itoper(8)}~/~{\bfseries\itshape itGUI(1)}
utilities.
The \verb|inttarg|'s \verb|oper()| call supports at least the
sub-functions enumerated in the {\bfseries\itshape itoper(8)}'s synopsis
(see \ref{inttarg.sw.itconfoper}). Generally speaking, the corresponding
operations have the essentially asynchronous nature, so, we implement
some kind of the finite state machine. This machine transits between well defined states
as a result of these operations execution. First of all, each operation should be
added by the \verb|oper()| into the FIFO queue (if a well defined operation
order permits it after the last already added operation). The queue is
implemented by the singly-linked tail queue \verb|STAILQ| (see
{\bfseries\itshape queue(3)}), and the operation is
represented in it by the \verb|struct op_entry| (see {\itshape toper.h}).
Each successfully added operation will be executed. Note that addition and
execution are performed simultaneously with {\bfseries\itshape mutex(9)} locking
arbitration. Execution is
% performed
completed in
% one or more (does not used now) stages each contain
two phases: execution itself (i.e. some work with CAMAC) and
asynchronous finish. The execution phase is performed by the \verb|oper()| (if
queue contains this command only), or by the kernel thread (after finishing
the previous
operation). The finish phase is made
% performed
by the kernel thread waked up by IRQ
handler, EoB or per-quantum {\bfseries\itshape callout(9)} handler, or
by own timed-out {\bfseries\itshape sleep(9)}.
% if thread slept with timeout which was expired.
Operation can have one or more repetitions. So, \verb|cycles #| operation
has \verb|#| repetitions, while the \verb|targon| one is continuous (up to the
\verb|targoff| operation appearing in the queue). The operation execution
and/or finishing failures lead to the queue discarding and the finite
state machine
appearing in the ``unknown'' (non-initialized) state.

The CAMAC hardware description and handling are separated from the
\verb|inttarg| module's source and grouped together in the single
header {\itshape inttarg\_hard\-wa\-re.h}~. This header
%In the current implementation the \verb|inttarg| module
uses macro interface {\bfseries\itshape kk(9)} specific for the KK009 crate
controller \cite{KK009} instead of the crate controller independent
interface {\bfseries\itshape camac(9)}, because the former
interface allows us to
slightly reduce the overhead for
% perform
each CAMAC cycle
% slightly faster
 \cite{IsupJINRC03}.

\subsubsection{Configuration {\bfseries\itshape itconf(8)} and control
{\bfseries\itshape itoper(8)} utilities}
\label{inttarg.sw.itconfoper}

\begin{verbatim}
itconf [-l] [-v] [-f<pflag>[,<pflag>...]] [-s{<arr_size>|-}]
   [-m<cnts_mask>] [-a] [-d{<driver>|-}] <module>
itconf -t [-l] [-v] <module>
\end{verbatim}

In the first synopsis form the {\itshape itconf} utility configures the
specified module \verb|<module>| (in our case usually \verb|inttarg|,
see \ref{inttarg.sw.inttarg4} and {\bfseries\itshape inttarg(4)}) for
work with driver \verb|kk0| by default and produces packets
with \verb/F_CRC|F_TIME/ (\verb|#define|d in the {\itshape packet.h})
flags by default.

In the second synopsis form the {\itshape itconf} utility tests the
configuration
of the specified module \verb|<module>| and writes it to the standard error
output.

The default behavior of {\itshape itconf} may be changed by the following
options:

\begin{description}
\item[{\ttfamily -d<driver>}]
Configure module for work with driver \verb|<driver>| instead of default
\verb|kk0|. \verb|-d-| means to use the compiled-in default for the driver.
The default driver name may be changed at {\itshape itconf} compile time.
\item[{\ttfamily -f<pflag>}] Set \verb|header.flag| field in the
\verb|make_pack()| produced packets in accordance to the \verb|<pflag>|
supplied. Valid values are: ``\verb|crc|'', ``\verb|time|'', ``\verb|none|''
(see {\bfseries\itshape packet(3)} for more details).
\item[{\ttfamily -s<arr\_size>}]
Set the size (number of members) of the arrays, which accumulate some
statistics during the burst, to \verb|<arr_size>|. Valid values are:
100..500 (corresponds to the burst of 1..5 sec). \verb|-s-| means to use
the compiled-in default for \verb|<arr_size>| (500), which can be
changed at {\itshape itconf} compile time.
\item[{\ttfamily -m<cnts\_mask>}]
Set the bit mask for the 10MSC counter regular inputs, which will be read
from CAMAC by configured module \verb|<module>|. The \verb|<cnts_mask>|
value means as follows: the 0th nonzero bit marks the 0th counter input to be
used, the 1st bit --- the 1st input, etc., up to the 7th bit for the 7th
input. \verb|-m| absence in the command string leads to using the
compiled-in default for \verb|<cnts_mask>| (\verb|0xff|, means --- to use all 8
inputs), which can be changed at {\itshape itconf} compile time.
\item[{\ttfamily -a}] Request to read ADC0..3 at each time quantum and produce
\verb|INTTARG_CYC_END2| data packets with corresponding data. By default (without
\verb|-a|) the ADC1 is still being read at EoB or Final Quant and reported at
\verb|INTTARG_CYC_BEG| packets.
\end{description}
%The {\itshape itconf} utility exits 0 on success, and $>0$ on error.

%|swstart
\begin{verbatim}
itoper [-l] [-v] [-b<#>] init|finish|start|stop|targon|targoff
    |status|cntcl|exec|done|clean|print
itoper [-N<#>] [-l] [-v] [-b<#>] targon
itoper -C<#> [-N<#>] [-l] [-v] [-b<#>] cycles
itoper -T<#> [-l] [-v] [-b<#>] chtarg
itoper [-r{<infile>|-}] [-l] [-v] [-b<#>] [-A<amax>] [-V<vmax>]
    [-c{-|<limsfile>}] settrj
itoper [-s{<outfile>|-}] [-l] [-v] [-b<#>] gettrj
\end{verbatim}
%itoper [-D<#>] [-l] [-v] [-b<#>] gotoinpos
%itoper [-F<#>] [-l] [-v] [-b<#>] test

In all the synopsis forms the {\itshape itoper} performs \verb|oper()| call
(see \cite{IsupJINRC10-35} and {\bfseries\itshape camacmod(9)}) with
sub-function \verb|fun| (see {\itshape inttarg.h}), defined by the first
supplied argument, on the CAMAC
module (usually {\bfseries\itshape inttarg(4)}) attached to the 0-th branch,
and writes the report about that action to the standard error output. The
\verb|init|, \verb|finish|, \verb|start|,
% \verb|swstart|,
\verb|stop|,
\verb|targon|, \verb|targoff|, \verb|cycles|, \verb|chtarg|, \verb|settrj|,
\verb|gettrj|, \verb|status|, and \verb|cntcl| are \verb|fun|s for production
usage and expected to have self-explained names.
Note with \verb|-v| the {\itshape itoper} also uses \verb|oper()| call with
\verb|status| sub-function.
The {\itshape itoper} may be used, for example, to implement some
commands in the supervisor configuration file
{\bfseries\itshape sv.conf(5)} (see \ref{inttarg.sw.itGUI}).

The default behavior of {\itshape itoper} may be changed, in particular,
by the following options:

\begin{description}
\item[{\ttfamily -C<\#>}] This mandatory flag supplies the number of cycles
\verb|<#>| to be serviced by the internal target before the implicit stop
(the third synopsis form of the {\itshape itoper}).
\item[{\ttfamily -N<\#>}] If this optional flag is supplied, the one (last)
of each \verb|<#>| cycles will not be serviced by the internal target --- so
called ``drop each N-th cycle'' mode (the
second and third synopsis forms of the {\itshape itoper}). This allows
another beam activity, for example, slow extraction. At the beginning
of each burst previous to the inactive one the packet of
\verb|INTTARG_INFO_0| type with \verb|WARN_PREP2DROP| value will be
generated.
\item[{\ttfamily -T<\#>}] This mandatory flag supplies the internal target
number \verb|<#>| to be made active (the fourth synopsis
form of the {\itshape itoper}). Valid values are 1..6.
\item[{\ttfamily -r<infile>}]
This optional flag supplies the \verb|<infile>| name of the input file which
contains the internal target trajectory to be programmed
(the fifth synopsis form of the {\itshape itoper}). \verb|-r-| means to use the
compiled-in default for input file name
({\itshape \$NGDPHOME/trj/in.trj}), which can be changed at
{\itshape itoper} compile time. The same is
used  if  the \verb|-r| is absent.
\item[{\ttfamily -s<outfile>}]
This optional flag supplies the \verb|<outfile>| name of the output file
where the current internal target trajectory should be stored
(the sixth synopsis form of the {\itshape itoper}). \verb|-s-| means to use
the compiled-in default for output file name
({\itshape \$NGDPHOME/trj/save.trj}), which can be changed at
{\itshape itoper} compile time.
\item[{\ttfamily -A<amax>}] Sets the acceleration upper limit
(in mksteps/msec$^2$) for the trajectory calculation to the supplied
\verb|<amax>| value. Default is 0.025.
\item[{\ttfamily -V<vmax>}] Sets the velocity upper limit (in mksteps/msec)
for the trajectory calculation to the supplied \verb|<vmax>| value. Default
is 5.0.
\item[{\ttfamily -c<limsfile>}]
Requires to read at startup the multipliers for the acceleration and velocity
limits from \verb|<limsfile>| (the  fifth  synopsis  form  of  the
{\itshape itoper}). \verb|-c-| means to use compiled-in default for
\verb|<limsfile>| ({\itshape \$NGDPHOME/etc/itGUI\_lims.cfg}), which can be changed
at {\itshape itoper} compile time. The same is
used if the \verb|-c| is absent. If the limits file opening or reading
fails, the multipliers are 1.0 for the whole time range (0..5000~msec).
\end{description}
%The {\itshape itoper} exits 0 on success, and nonzero on error.

The \verb|<infile>| and \verb|<outfile>| contain the pair of ASCII float
numbers delimited by space and/or tab symbol(s) per each line. (Lines are
delimited by the newline symbol as it is usual for UNIX
textual files). The first number is the time in msec (abscissa), the
second --- a rotating angle in arc degrees (ordinate). Lines with comment symbol
``\verb|#|'' in the first position are ignored. For example,
% by default the {\itshape \$NGDPHOME/trj/in.trj} contains the following
the requested trajectory in Fig.~\ref{inttarg.fig.trj_canvas}
is as follows:\\[-8mm]
%\begin{verbatim}
%#time   position (degrees)
%0       0
%900     24.3063
%3100    32.3
%5000    0
%\end{verbatim}
\begin{verbatim}
#time   position (degrees)
0       0
900     33.2
3300    35.0
4300    0
\end{verbatim}

Each line of the \verb|<limsfile>| contains the four fields
delimited by space and/or tab symbol(s). (Lines are
delimited by a newline symbol as it is usual for UNIX
textual files). The first field is a keyword, the \verb|alim| means the line
belongs to the acceleration's limitation, the \verb|vlim| --- to the velocity's one.
The fourth field is an ASCII float and represents the multiplier for the limit,
which is defined by the \verb|-A|/\verb|-V| option or by default. The
second and third fields are ASCII integers from 0 to 499 and represent the
lower and upper boundaries of the trajectory range (in the 10~msec units),
where the multiplier will be applied. Lines with comment symbol
``\verb|#|'' in the first position are ignored. For example,
% by default the
%{\itshape \$NGDPHOME/etc/itGUI\_lims.cfg} contains something like the
%following:
the requested trajectory in Fig.~\ref{inttarg.fig.trj_canvas} uses the
following limit multipliers:\\[-8mm]
%\begin{verbatim}
%#type	min	max	mult
%alim	0	50	5.0
%alim	50	100	0.2
%vlim	0	500	1.0
%\end{verbatim}
\begin{verbatim}
#type   min     max     mult
alim    0       50      7.0
alim    50      200     0.3
vlim    0       150     1.2
\end{verbatim}

\subsubsection{The {\bfseries\itshape itGUI(1)} user control utility}
\label{inttarg.sw.itGUI}

%|swstart
\begin{verbatim}
itGUI [-l] [-v] [-b<#>] [-a] [-L] [-M<mask>|-m<mask>] [-A<amax>]
   [-V<vmax>] [-r{infile|-}] [-s{outfile|-}] [-p{-|<pidfile>}]
   [-c{-|<limsfile>}]
itGUI -o [-l] [-v] [-b<#>] init|finish|start|stop|targon
   |targoff|status|cntcl|exec|done|clean|print
itGUI -o [-N<#>] [-l] [-v] [-b<#>] targon
itGUI -o -C<#> [-N<#>] [-l] [-v] [-b<#>] cycles
itGUI -o -T<#> [-l] [-v] [-b<#>] chtarg
itGUI -o [-r{<infile>|-}] [-l] [-v] [-b<#>] [-A<amax>] [-V<vmax>]
   [-c{-|<limsfile>}] settrj
itGUI -o [-s{<outfile>|-}] [-l] [-v] [-b<#>] gettrj
\end{verbatim}
%itGUI -o [-D<#>] [-l] [-v] [-b<#>] gotoinpos
%itGUI -o [-F<#>] [-l] [-v] [-b<#>] test

\begin{figure}[htb]
\epsfig{width=\textwidth
,bbllx=0pt,bblly=0pt,bburx=576pt,bbury=432pt
,file=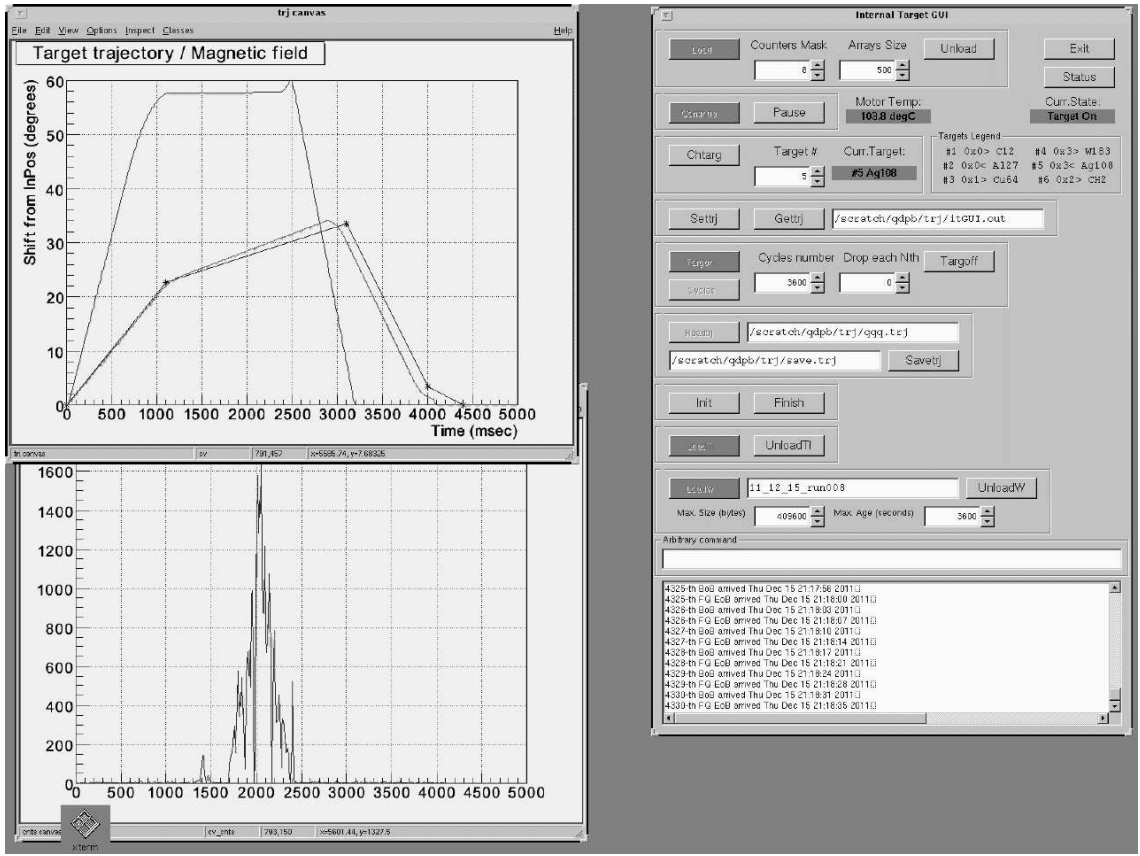}
\caption{The {\bfseries\itshape itGUI(1)} screenshot. See text for description.}
\label{inttarg.fig.itGUI}
\end{figure}

\begin{figure}[htb]
\epsfig{width=\textwidth
,bbllx=0pt,bblly=0pt,bburx=567pt,bbury=422pt
,file=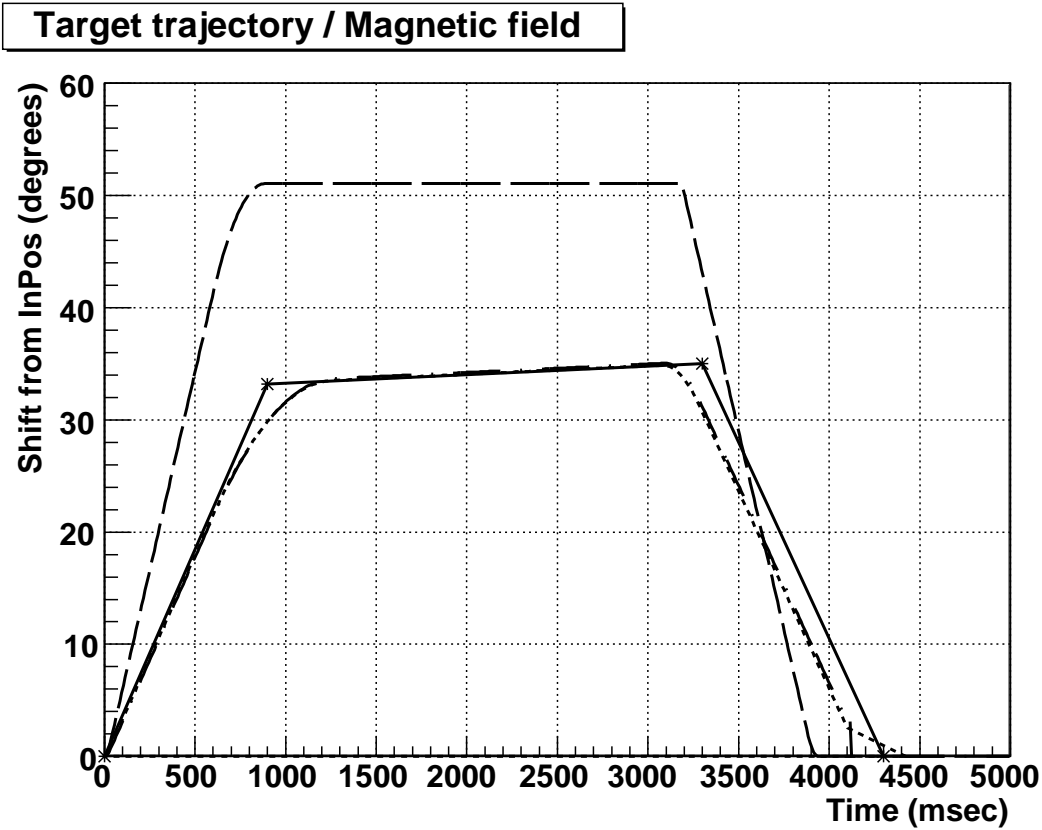}
\caption{The {\bfseries\itshape itGUI(1)} trajectories window. See text for description.}
\label{inttarg.fig.trj_canvas}
\end{figure}

\begin{figure}[htb]
\epsfig{width=\textwidth
%,bbllx=0pt,bblly=0pt,bburx=567pt,bbury=422pt
,file=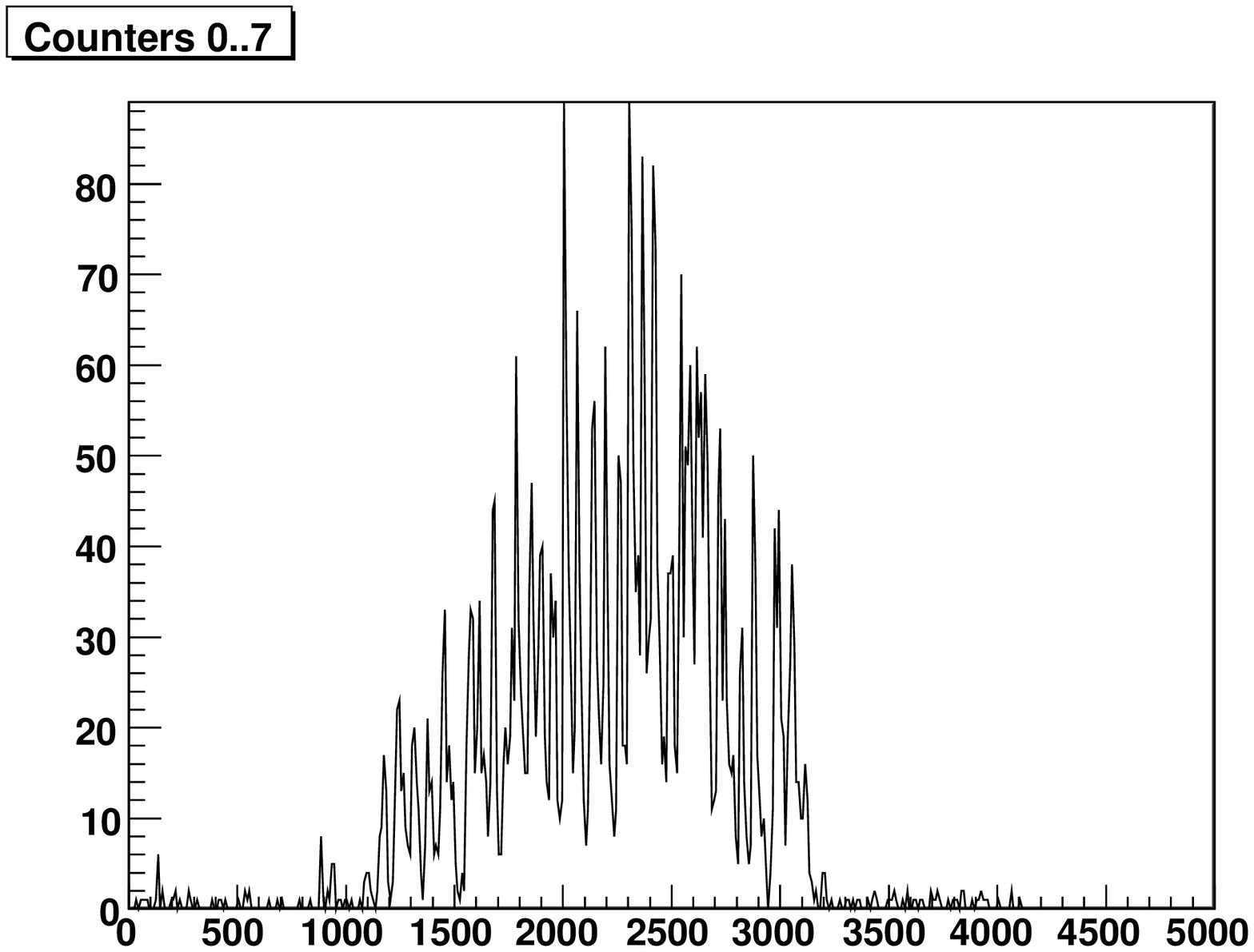}
\caption{The {\bfseries\itshape itGUI(1)} counters window. See text for description.}
\label{inttarg.fig.cnts_canvas}
\end{figure}

The {\itshape itGUI} provides the graphic user interface (GUI) for
conversation with the IntTarg
CDAQ system as well as for the graphic representation of the read out experimental
data using the ROOT framework \cite{ROOTproc} libraries. In the first synopsis
form the {\itshape itGUI} draws one main window of buttons
% (see
%% section MAIN WINDOW
%\ref{inttarg.sw.itGUI.mainwin})
(in Fig.~\ref{inttarg.fig.itGUI} -- the
% upper
right window ``{\sffamily
Internal Target GUI}'')
and some additional windows
% (see
%% section DATA WINDOWS
%\ref{inttarg.sw.itGUI.datawins})
to display target trajectories (the upper left window) and
other acquired data: magnetic field (the same window), multiscaler input(s)
(the lower left window) and ADC
channels (not shown). After that the {\itshape itGUI}
tests the current IntTarg CDAQ state to highlight buttons, correspondingly,
launches the
child process
% thread (\verb|TThread|)
to read {\itshape ngdp} packets from the standard input (usually supplied by
the {\bfseries\itshape ngget(1)}),
and goes into the endless loop (\verb|TApplication::Run()|) of the graphic
events handling. Note the {\itshape itGUI} could be safely and consistently
restarted at any time during the IntTarg CDAQ working without the latter state
changes.

If the {\itshape itGUI} called with the \verb|-o| flag or under the
{\itshape itoper} name, it behaves the same way as\footnote{
The {\bfseries\itshape itGUI(1)} shares the corresponding source with the
{\bfseries\itshape itoper(8)}.
} {\bfseries\itshape itoper(8)}
utility in the corresponding synopsis forms (without \verb|-o|, of course),
see
% itoper(8) manual pages
\ref{inttarg.sw.itconfoper}.

The default behavior of {\itshape itGUI} may be changed, in particular,
by the following options\footnote{
Options already described in \ref{inttarg.sw.itconfoper}
are not mentioned here. For the {\ttfamily <infile>}/{\ttfamily <outfile>}
and {\ttfamily <limsfile>} formats see \ref{inttarg.sw.itconfoper}.
}:

\begin{description}
\item[{\ttfamily -a}] Handle ADC0..3 histograms instead of ADC1
channel only for the stepper
motor temperature calculation by default.
\item[{\ttfamily -L}] Leave mode --- to survive after the data reading child
process termination (usually after EOF obtaining while the offline data file
reading). It allows us to play with the graphic output without future redraws.
\item[{\ttfamily -M<mask>}] Supplies the 8-bit mask \verb|<mask>|, whose bits
correspond to 10MSC regular inputs. Each nonzero bit means the corresponding
counter to be additive during the current run instead of the counter
resetting per each cycle by default.
\item[{\ttfamily -m<mask>}] The same as \verb|-M|, however, it normalizes
additive counters to the packet's number.
\end{description}
%The {\itshape itGUI} exits 0 on success, and nonzero on error.

The {\itshape itGUI} is implemented with having in mind
the supervisor utility concept (see \cite{IsupJINRC01-116}). According to
this concept the {\itshape itGUI} has the configuration file
(named by default {\itshape \$NGDPHOME/etc/inttargsv.conf}) in the
{\bfseries\itshape sv.conf(5)} format (really Makefile, see also
{\bfseries\itshape make(1)}). This file establishes the correspondence between
the user commands (``targets'' in {\bfseries\itshape make(1)} terminology) and
actions which should be performed.
% at its execution.
This textual file could be
revised easily without the {\itshape itGUI} recompile.

%\centerline{\bfseries {\bfseries\itshape itGUI(1)} main window}
%%\label{inttarg.sw.itGUI.mainwin}

The main {\bfseries\itshape itGUI(1)} window (\verb|TGMainFrame|) contains (see
Fig.~\ref{inttarg.fig.itGUI}) the buttons
(\verb|TGTextButton|), the string (\verb|TGTextEntry|) and number
(\verb|TGNumberEntry|) input fields, the current
state and target indication fields (\verb|TGLabel|), and debug output viewer
(\verb|TGTextView|). Each button could be pressed by the left mouse button
single-click, as well as the input focus could be placed into the input
fields. The generic system startup direction is from up to down (and
system stopping --- from down to up). The buttons with ``On'' and ``Off''
meanings are placed in the same horizontal ``engraved'' frame from the left
to the right.

Each button could be in the following states:\\
$\bullet$ active (could be pressed, black foreground);\\
$\bullet$ pressed (reverse shadow, grey foreground); or\\
$\bullet$ inactive (could not be pressed, grey foreground).

Each button could display the following operation states:\\
$\bullet$ the operation could be tried to perform (grey background);\\
$\bullet$ the operation successfully done (green background);\\
$\bullet$ the operation in progress or in queue (yellow background);\\
$\bullet$ the operation failed to be added to the queue or to be done (red
background).

The buttons
% (each but \verb|Exit| corresponds to command (target) with the
%same name without capitalization from the supervisor configuration file)
are:\\[-8mm]
\begin{description}
\item[{\ttfamily Load}] loads and configures the
{\bfseries\itshape inttarg(4)} kernel module.
The corresponding number input fields: \verb|Counters Mask| --- mask of the
10MSC input(s) to be read (0..\verb|0xff|), \verb|Arrays Size| --- base
array's size (500 per 5 second accelerator burst).
\item[{\ttfamily Unload}] (counterpart of previous) unloads the
{\bfseries\itshape inttarg(4)} kernel module.
\item[{\ttfamily Exit}] sends \verb|SIGTERM| to the {\itshape itGUI}'s process
group and exits the {\itshape itGUI}. Note the {\itshape itGUI}'s termination
does not change the IntTarg CDAQ state in any way.
\item[{\ttfamily Status}] collects the status outputs from the number of system
parts and displays these outputs in the debug output viewer.
\item[{\ttfamily Readtrj}] reads the requested trajectory from the input file
under the name given
% entered into
in the string input field to the right from the \verb|Readtrj|
button. This reading totally replaces the current requested trajectory, as
well as its interactive
modification%
%, see
%% section DATA WINDOWS
%\ref{inttarg.sw.itGUI.datawins}%
, and leads to the calculated trajectory updating.
\item[{\ttfamily Savetrj}] saves the current requested trajectory in the
output file under the name given in
% entered into
the string input field to the left from the
\verb|Savetrj| button.
\item[{\ttfamily Continue}] starts the data acquisition, i.e. handling of
the BoB interrupts. In this state the system is ready for target walking, and
the {\bfseries\itshape targinfo(1)} server has the data to be distributed.
\item[{\ttfamily Pause}] (counterpart of previous) stops the BoB interrupts
handling. In this state the ITS
% internal target
hardware could be safely powered off/on.
\item[{\ttfamily Settrj}] supplies the current calculated trajectory for
the {\bfseries\itshape inttarg(4)} kernel module.
\item[{\ttfamily Gettrj}] gets the calculated trajectory from the
{\bfseries\itshape inttarg(4)} kernel module and prints it in the debug file
under the name given in
% entered into
the string input field to the right from the \verb|Gettrj|
button.
\item[{\ttfamily Chtarg}] sets the target with number \verb|<#>| to be a
current (ready to walk) target. The corresponding number input field:
\verb|Target #| --- the number of the target to be the current one (valid
numbers are 1..6). The
correspondences between the numbers and materials are displayed to the
right from the \verb|Chtarg| group frame.
\item[{\ttfamily Targon}] allows the current target to walk during each cycle
infinitely (up to explicit denying by the \verb|Targoff| pressing). The target
starts to walk at the nearest BoB. During the latest of
each \verb|N| bursts the target could be not walking, if the
% nonzero
\verb|N|~$\ge 2$ value is supplied with the corresponding number input field
\verb|Drop each Nth|. Valid numbers are 2..3600, and 0, 1, which mean that
the target does not drop the bursts.
\item[{\ttfamily Targoff}] (counterpart of previous) denies the current target
to walk. The target stops to walk after the nearest EoB or before
the nearest BoB.
\item[{\ttfamily Cycles}] allows the current target to walk during each
cycle of the nearest \verb|#| cycles (or up to explicit denying by the
\verb|Targoff| pressing), supplied with the corresponding number input field
\verb|Cycles number| (valid numbers are 1..3600). The target starts to
walk at the nearest BoB. During the latest of each \verb|N|
bursts the target could be not walking, if the
% nonzero
\verb|N|~$\ge 2$ value
is supplied with the corresponding number input field \verb|Drop each Nth|.
Valid numbers are 2..3600, and 0, 1, which mean that
the target does not drop the bursts.
\item[{\ttfamily Init}] initializes ITS
% internal target
hardware, in particular,
CAMAC modules in the read-out crate, and positions the target into
the nearest InPos. Note the \verb|Init| is also a part of \verb|Load|.
\item[{\ttfamily Finish}] (counterpart of previous) de-initializes the internal
target hardware. In the current design
% Evidently/possibly
it is not needed at all.
\item[{\ttfamily LoadW}] loads the {\bfseries\itshape writer(1)} --- utility
to write the packet stream into files on HDD. The corresponding input fields:\\
(string) --- base name of the data files for the current {\bfseries\itshape writer(1)}
run;\\
\verb|Max. Size (bytes)| (number) --- the recommended size for each file;\\
\verb|Max. Age (seconds)| (number) --- the recommended age for each file.
\item[{\ttfamily UnloadW}] (counterpart of previous) unloads the
{\bfseries\itshape writer(1)} utility and consequently terminates the data
writing of the current run.
\item[{\ttfamily LoadTI}] loads the {\bfseries\itshape targinfo(1)} ---
server, which distributes the read
out internal target trajectory to its already registered clients.
\item[{\ttfamily UnloadTI}] (counterpart of previous) unloads the
{\bfseries\itshape targinfo(1)} server.
\end{description}

Each but \verb|Exit|, \verb|Readtrj|, \verb|Savetrj| button corresponds to the
command (target) under the
same name (without capitalization) in the supervisor configuration file, so
the IntTarg CDAQ control has two functionally equivalent interfaces: graphic
(by {\bfseries\itshape itGUI(1)}) and textual (by configuration file
{\bfseries\itshape make(1)}ing).
Note, however, that {\bfseries\itshape itGUI(1)} itself uses the configuration
file to perform the complex commands (\verb|[un]load|, \verb|[un]loadw|,
\verb|[un]loadti|, \verb|status|) only. In contrast, the simple commands
(see \verb|-o| synopsis forms) corresponding directly to
{\bfseries\itshape inttarg(4)}'s \verb|oper()| sub-functions, are performed
internally using the code shared with {\bfseries\itshape itoper(8)}.

The indication fields in the main window are:\\[-8mm]
\begin{description}
\item[{\ttfamily Curr.State:}]
(under the \verb|Status| button) displays the current state of the
operations queue (see \ref{inttarg.sw.inttarg4}), one of: `Unknown'
(on grey background --- before \verb|Load|
and after \verb|Unload|, on red --- otherwise), `Init', `Start', `Stop',
`Cycles', `Target On' (all on green).
\item[{\ttfamily Curr.Target:}]
(to the right for the \verb|Chtarg| button) displays the current target number
and material (on the green background --- after successful target
change, on the yellow one --- after startup and while the target changing,
on the red --- after the unexpected target change or after obtaining the
invalid number of the current target), for example: `\verb|#5 Ag108|'.
\end{description}

The string input field named ``{\sffamily Arbitrary command}'' (just above
the debug output viewer) allows the user to perform any target from the
supervisor configuration file.

%\vspace*{2mm}

%\centerline{\bfseries {\bfseries\itshape itGUI(1)} data windows}
%%\label{inttarg.sw.itGUI.datawins}

%\vspace*{-3mm}

The {\bfseries\itshape itGUI(1)} data windows are as follow:\\[-8mm]
\begin{description}
\item[trj canvas] The internal target trajectories --- requested (in black,
with asterisk points), calculated (in red) and read out (in green, only after
cycles with the active target) from the 8th 10MSC up/down
input --- as well as the magnetic field (in blue) read out from the
9th 10MSC up/down input are displayed in the single \verb|TCanvas| named
``{\sffamily trj canvas}''. On the {\itshape itGUI} screenshot
(Fig.~\ref{inttarg.fig.itGUI}) we can see this canvas as the left upper window
frame with the ``{\sffamily Target trajectory / Magnetic field}'' title.
With the higher resolution such canvas is shown in
Fig.~\ref{inttarg.fig.trj_canvas}. The calculated (dotted) and read out
(dash-dotted) trajectories are
 in some segments below
% near
the requested trajectory (the solid
line with two asterisks) and approximately coincide. The last segment of the
read-out curve goes to zero vertically, because the multiscaler is not read after
the final time quantum, however, the target really goes into InPos with the
fixed 1 mkstep/msec velocity, as it is shown by the calculated curve.
The arbitrary scaled magnetic field
%\footnote{
%Unfortunately we had not the down input for it from accelerator electronics.
%}
(dashed curve) is above all the others.
The abscissa is the time in
milliseconds, the ordinate is a target shift in arc degrees (InPos at $0^{\circ}$,
beam center approximately at $30^{\circ}$). Note, the requested and calculated
trajectories could not be very close to each other (as in
Fig.~\ref{inttarg.fig.trj_canvas}),
because the trajectory calculation algorithm has the upper limits for the
target velocity (leads to the lower slope angle) and acceleration (leads the to
angle smoothing).
% and target shift ``prolongation'').
To be more flexible, these limits could
be varied along the trajectory according to the configuration file
{\itshape itGUI\_lims.cfg} (format described in \ref{inttarg.sw.itconfoper}).
So, the trajectory in
Fig.~\ref{inttarg.fig.trj_canvas}
is calculated with the following limits:\\
\verb|amax|$ < 0.175$~mksteps/msec$^2$ at 0..500~msec\\
\verb|amax|$ < 0.0075$~mksteps/msec$^2$ at 500..2000~msec\\
\verb|vmax|$ < 6.0$~mksteps/msec at 0..1500~msec\\
(in other time ranges the both limits are default).
The requested trajectory could be read from the input file under the name
given in
%entered into
the corresponding \verb|TGTextEntry| by pressing the
%\verb|TGTextButton|
\verb|Readtrj| (see
%% section MAIN WINDOW
%\ref{inttarg.sw.itGUI.mainwin}
above). Reading from
the file totally replaces the current requested trajectory. The
current requested trajectory could be interactively modified. The
point could be added by the left mouse button double-click, while
the existing point could be removed by the middle mouse button double-click
(see also the hot keys below). Note, neither the points
nor the whole trajectory should be moved. After each modification of
the requested trajectory the calculated trajectory is updated. The current
requested trajectory could be saved in the output
file under the name given in
% entered into
the corresponding \verb|TGTextEntry| by
pressing the \verb|Savetrj|
% \verb|TGTextButton|
(see
%% section MAIN WINDOW
%\ref{inttarg.sw.itGUI.mainwin}
above). The following hot keys are supported while mouse focus is in
% this window:\\
``{\sffamily trj canvas}'':\\
`a', `A' --- add the point under the current mouse position;\\
`d', `D' --- delete the point under the current mouse position;\\
`i', `I' --- input the point using the \verb|TGInputDialog| (the coordinates
should be entered as the time-angle pair of the float numbers delimited by
space and/or comma);\\
`u', `U' --- remove
% undo
the last point added by the left double-click, `A', `a', `I',
`i' (could be used many times);\\
`p', `P' --- print the calculated trajectory into the debug file under the
name given in
% entered into
the \verb|TGTextEntry| to the right from the \verb|Gettrj|
%\verb|TGTextButton|
(see
%% section MAIN WINDOW
%\ref{inttarg.sw.itGUI.mainwin}
above);\\
`r', `R' --- read out the requested trajectory from the input file under the
name given in
% entered into
the \verb|TGTextEntry| to the right from the \verb|Readtrj|
%\verb|TGTextButton|
(see
%% section MAIN WINDOW
%\ref{inttarg.sw.itGUI.mainwin}
above);\\
`c', `C' --- clean out the trajectory (only (0,0) and (5000,0) points are
preserved);\\
`s', `S' --- save the current requested trajectory in the output file
under the name given in
% entered into
the \verb|TGTextEntry| to the left from the \verb|Savetrj|
%\verb|TGTextButton|
(see
%% section MAIN WINDOW
%\ref{inttarg.sw.itGUI.mainwin}
above);\\
`T' --- supplies the current calculated trajectory for the
{\bfseries\itshape inttarg(4)} kernel module (the same as \verb|Settrj|%
%\ \verb|TGTextButton|%
, see
%% section MAIN WINDOW
%\ref{inttarg.sw.itGUI.mainwin}
above);\\
`t' --- gets the calculated trajectory from the {\bfseries\itshape inttarg(4)}
kernel module and prints it in the debug file under the name given in
% entered into
the \verb|TGTextEntry| to the right from the \verb|Gettrj|
% \verb|TGTextButton|
(see
%% section MAIN WINDOW
%\ref{inttarg.sw.itGUI.mainwin}
above);\\
`q', `Q' --- quit the {\itshape itGUI} (the same as \verb|Exit| button pressing%
%\ \verb|TGTextButton|%
, see
%% section MAIN WINDOW
%\ref{inttarg.sw.itGUI.mainwin}
above).
\item[cnts canvas] From 0 up to 8 counter values, read-out from the 0..7th
10MSC inputs, are represented by
% \verb|TGraph|s
\verb|TH1D|s and displayed in the single
\verb|TCanvas| ``{\sffamily cnts canvas}'' iconified at startup.
In Fig.~\ref{inttarg.fig.itGUI} it is the lower left window frame with
% the ``{\sffamily Graph}''\footnote{
%In the current {\bfseries\itshape itGUI(1)} version we have ``{\sffamily
%Counters 0..7}''.
%} string and
the single curve, which represents the interaction intensity monitor counts
for the last processed cycle.
%, see chapter~\ref{inttarg.run_exp} for more details.
This plot allows the user to tune the target movement trajectory.
Numbers and positions of the inputs to be used are configured (see
{\bfseries\itshape itconf(8)}) while loading of the
{\bfseries\itshape inttarg(4)} kernel module.
The intensity monitor counts during the single cycle for the C target
run is shown in Fig.~\ref{inttarg.fig.cnts_canvas}.
\item[cv\_adc\_N] Four ADC channel histograms (\verb|TH1|) are displayed each
in its own window (\verb|TCanvas|)
``{\sffamily cv\_adc\_0}''..``{\sffamily cv\_adc\_3}'' iconified at
startup (in Fig.~\ref{inttarg.fig.itGUI} we can't see them), if the
\verb|-a| option is specified.
\end{description}

\subsubsection{The {\bfseries\itshape targinfo(1)} trajectory server}
\label{inttarg.sw.targinfo}

\hspace*{4mm}
The {\bfseries\itshape targinfo(1)} is a server, which
provides its clients with the internal
target trajectory data in each accelerator cycle at the end of this cycle.

\verb/targinfo [-l] [-t] [-f<#>] [-p{-|<pidfile>}] [-a<address>[ ...]]/

In this synopsis form the {\itshape targinfo} listens to the TCP/IP
socket on the host {\ttfamily \IThost}, port 12345 to get client
connection requests,
% (up to \verb|CLIENTS_MAX| clients simultaneously, usually 5),
and reads packets from the standard input.
% (usually supplied by the {\bfseries\itshape ngget(1)}).
From each of the obtained \verb|INTTARG_CYC_BEG| packets
the {\itshape targinfo} collects the BoB
%begin of accelerator burst
timestamp, while from each of the \verb|INTTARG_CYC_END| ones it collects
the microstep number array of the ITS
% internal target
stepper motor. Once per cycle the {\itshape targinfo} writes the cycle
timestamp and some target trajectory data (format
described below) to all the currently connected clients (if any).
Up to \verb|CLIENTS_MAX| clients (usually 5) can be serviced 
simultaneously, connections closed by the peer are recycled.

\begin{figure}[htb]
\epsfig{width=\textwidth
,file=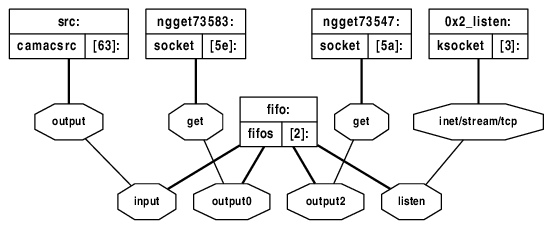}
\caption{
The IntTarg CDAQ core is implemented by the {\itshape ngdp} graph.
}
\centerline{\small Rectangles are nodes with: name (up), type (left), ID (right);}
\centerline{\small octagons are hooks named within.}
\centerline{\small {\ttfamily ng\_fifos} has two local output streams
  through {\ttfamily ng\_socket}s,}
\centerline{\small as well as {\ttfamily listen()}ing {\ttfamily ng\_ksocket}.}
\label{inttarg.fig.graph}
\end{figure}

\begin{figure}[htb]
\epsfig{width=\textwidth
,bbllx=10pt,bblly=36pt,bburx=585pt,bbury=404pt
,file=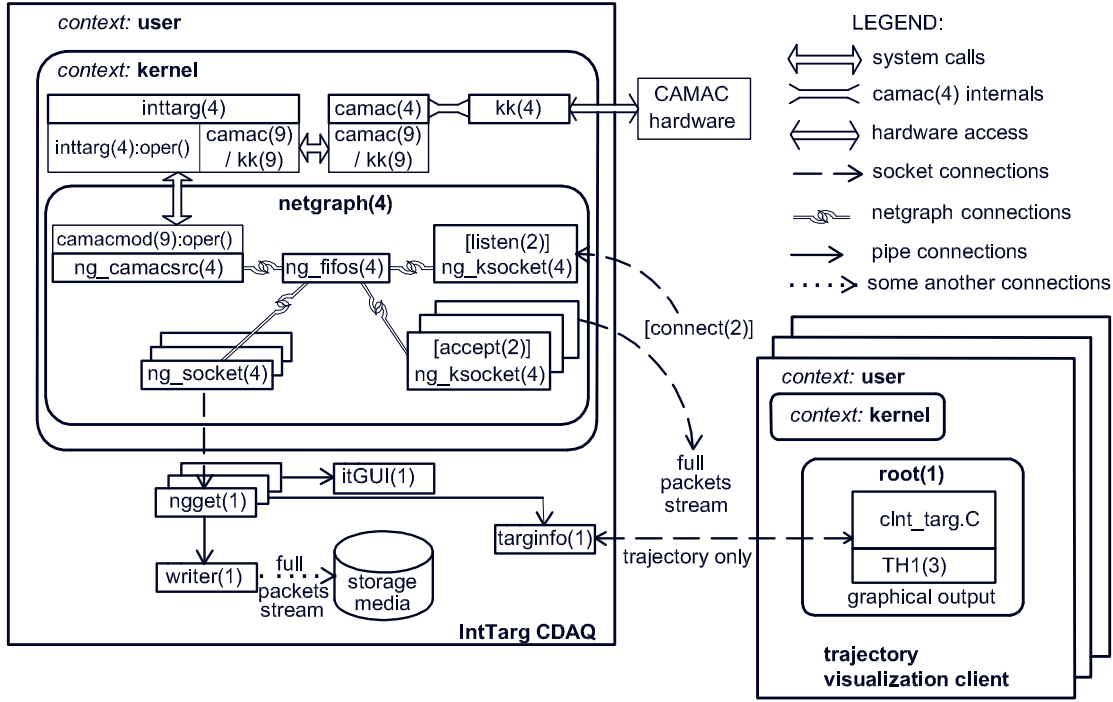}
\caption{Overall IntTarg CDAQ layout.}
\label{inttarg.fig.schema}
\end{figure}

The default behavior of the {\itshape targinfo} may be changed by the following
options:\\[-8mm]
\begin{description}
\item[{\ttfamily -t}] Exits at all negative conditions from
{\bfseries\itshape packet(3)} functions
instead of the exit only at $-$3 by default.
\item[{\ttfamily -f<\#>}] Assigns the supplied \verb|<#>| number to the
\verb|pack_flags| variable for the
{\bfseries\itshape packet(3)} \verb|read_pack()| function.
\verb|-f| absence in the command string leads to using the compiled-in default
for \verb|<#>| (\verb|F_CRC| (see {\itshape packet.h}), because this is the only
checkable value for reading).
\item[{\ttfamily -a<address>}] Restricts clients to connect from specified
\verb|<address>| only instead of allowing the client to connect from any one
by default. \verb|<address>| can be an
Internet name in domain notation or IP address as four decimals
separated by dots. Note that \verb|-a| option can be specified with
different \verb|<address>|es up to \verb|CLIENTS_MAX| times.
\end{description}
%The {\itshape targinfo} utility exits 0 on success, and $>0$ on error.

The data format preserved after workaround implementation under the DAQ system
of the SCAN setup \cite{AfanPTE08} is as follows:\\
$\bullet$ timestamp of the burst begin (8 bytes of the
\verb|struct timeval|);\\
$\bullet$ signed target shift in 1/10 mm per each 20 msec (the 250
\verb|int16_t|s allows us to cover up to 5 second burst).

The format also indicates the following situations:\\
$\bullet$ A nonactive target. If the target was not injected into the beam
during the current cycle, all the 250 \verb|int16_t|s are equal to 0.\\
$\bullet$ Some erroneous target behaviour. If the target was walking
incorrectly, all 250 \verb|int16_t|s are equal to \verb|SHRT_MAX|
(\verb|0x7fff|).\\
$\bullet$ The next cycle will be nonactive. The ITS
%internal target
control supports a special operation mode, in
which the target remains inactive each N-th of N cycles to allow other
beam activities. In this mode, if the trajectory data are \verb|SHRT_MAX-1|
(\verb|0x7ffe|), the internal target will be inactive during the next
cycle. Note, these data will be written soon after obtaining
\verb|INTTARG_CYC_BEG| packet
in contrast with all the other data types generated at
\verb|INTTARG_CYC_END| arrival. Note also, that the normal trajectory
output will
be generated for the current cycle, too, and during the next (inactive
for the internal target) cycle the trajectory data will be zero (as it
should be expected).
%(250 \verb|int16_t|s are equal to
%\verb|0x7ffe|). Issued only in the ``drop one of N-th'' mode soon after begin
%of previous active burst in addition to the usual trajectory data.

\subsection{Bringing all things together}
\label{inttarg.sw.startup}

\hspace*{4mm}
The {\itshape ngdp} graph used by the IntTarg CDAQ is shown in
Fig.~\ref{inttarg.fig.graph}. It is very much like {\itshape ngdp}'s
CAMAC Front-End Modules (FEM) level proposed in \cite{IsupJINRC10-34}. The
node named \verb|fifo:| of type \verb|ng_fifos| is instantiated
(\verb|mkpeer|ed) at the {\ttfamily \IThost} host bootstrap time
using the script {\itshape \$NGDPHOME/etc/fifo.ngctl} processed by the
{\bfseries\itshape ngctl(1)}.
% by the {\bfseries\itshape ngctl(1)} which processes script
%{\itshape \$NGDPHOME/etc/fifo.ngctl}.
During \verb|ng_fifos| startup it
\verb|mkpeer|s the node of type \verb|ng_ksocket| with automatically chosen
name (\verb|0x2_listen:| in Fig.~\ref{inttarg.fig.graph}), and \verb|connect|s
the remote hook \verb|inet/stream/tcp| with its own hook \verb|listen|. So, the
\verb|ng_fifos| and its \verb|listen()|ing \verb|ng_ksocket|
are still ready from OS's boot to shutdown.

The \verb|ng_fifos| provides identical packet streams through all the
currently connected outputs, both the local and remote ones.
%??? In
%Fig.~\ref{inttarg.fig.graph}
%\verb|ng_fifos| has two local output streams
%through \verb|ng_socket|s,
%as well as \verb|listen()|ing \verb|ng_ksocket|.
Each output could be
connected and disconnected without disturbing other outputs, so the
packet stream consumers ({\bfseries\itshape writer(1)},
{\bfseries\itshape targinfo(1)}, and {\bfseries\itshape itGUI(1)}) could be
started and terminated independently of each other. These utilities
are launched by the \verb|loadw|, \verb|loadti| and \verb|loadgui| commands
from the {\itshape \$NGDPHOME/etc/inttargsv.conf} to be readers of the pipes,
where the writers are the {\bfseries\itshape ngget(1)}s. Each
{\bfseries\itshape ngget(1)} \verb|mkpeer|s the \verb|ng_socket| instance
(\verb|ngget73583:| and \verb|ngget73547:| in Fig.~\ref{inttarg.fig.graph})
and \verb|connect|s it by hook \verb|get| with \verb|ng_fifos|'s hook
\verb|output<N>|. The simultaneously allowed numbers of both the
\verb|output<N>| hooks and \verb|accept()|ing \verb|ng_ksocket|s are the
compiled-in parameters of \verb|ng_fifos|.

The \verb|load| command of the supervisor configuration file, in particular,
loads the {\bfseries\itshape inttarg(4)} interrupt handler and executes
the {\bfseries\itshape ngctl(1)} utility to
proceed the script {\itshape \$NGDPHOME/etc/camacsrc.ngctl}, which
\verb|mkpeer|s the node named \verb|src:| of type \verb|ng_camacsrc|. This node
communicates (see \cite{IsupJINRC10-35} for details) with CAMAC kernel
module {\bfseries\itshape inttarg(4)}, and \verb|connect|s its own hook
\verb|output| with the hook \verb|input| of the \verb|ng_fifos|. After that the
IntTarg CDAQ graph has all components which are
% possible
%available in
provided by the current design.

The supervisor configuration file
{\itshape \$NGDPHOME/etc/inttargsv.conf}
allows the user to control the IntTarg
CDAQ in the command-line mode through a simple textual terminal (without GUI).
Of course, the requested trajectory could not be corrected interactively in
this mode, and user can't see all the read-out data. However, the trajectory
file is textual (see \ref{inttarg.sw.itconfoper}), so it could be edited easily.

The overall IntTarg CDAQ layout is pictured in
Fig.~\ref{inttarg.fig.schema}, where we can see the host {\ttfamily \IThost}
as the rectangle entitled ``{\sffamily IntTarg CDAQ}''. In the user context the three
processes ({\bfseries\itshape itGUI(1)}, {\bfseries\itshape targinfo(1)} and
{\bfseries\itshape writer(1)}) obtain three identical packet streams from
three {\bfseries\itshape ngget(1)}s, which read three
{\bfseries\itshape ng\_socket(4)}s connected to {\bfseries\itshape ng\_fifos(4)}.
The packet streams with the same contents could be also transferred remotely
through the {\bfseries\itshape accept(2)}ing
{\bfseries\itshape ng\_ksocket(4)}s instantiated after client's
{\bfseries\itshape connect(2)}ion to the {\bfseries\itshape listen(2)}ing
{\bfseries\itshape ng\_ksocket(4)}. (The {\bfseries\itshape netgraph(4)}
behaviour mimics the BSD socket handling scheme.) In the present state we
have no remote consumers of the full packet stream, however, they can
appear in future (see \ref{inttarg.sw.upgrades}).

The rectangle ``{\sffamily trajectory visualization client}'' in
Fig.~\ref{inttarg.fig.schema} is another host which executes one of the
possible clients of the {\bfseries\itshape targinfo(1)} server --- the ROOT
script
{\itshape clnt\_targ.C}~. Note the {\bfseries\itshape targinfo(1)} distributes
the read out trajectory only, not the packet stream (see also
\ref{inttarg.sw.targinfo}). So, the {\bfseries\itshape targinfo(1)} is
preserved mostly for the backward compatibility and could be retired in
future.

\subsection{Nuclotron run experience and future directions}
\label{inttarg.sw.upgrades}

\hspace*{4mm}
The IntTarg CDAQ was used to control 
% ITS
the internal target during
the March and December 2011 Nuclotron accelerator runs.
%for DELTA \cite{krasnov2,delton} and DSS \cite{dss,piyadin,gurchin,dss-2011}
%experiments data taking.
The total beam time of the ITS
% internal target
operation was
approximately 150 hours on the deuteron beam at $T_{kin} = 250$--500
MeV/nucleon.
In particular, the ``drop each N-th cycle'' mode was successfully used
in March 2011.

The IntTarg CDAQ usage experience during the March 2011 Nuclotron
accelerator run provides
some hints to improve the software, first of all
{\bfseries\itshape itGUI(1)}, in some aspects.
So, the ROOT \verb|TGraph| was replaced by the \verb|TH1D| to represent most of
the visualized curves with many (some hundreds) points since the \verb|TGraph|
has shown a dramatically low visualization performance (ROOT 5.16).
Also the ROOT \verb|TThread|
% and POSIX {\bfseries\itshape pthread(3)}
was eliminated in favor of traditional UNIX child process \verb|fork()|ing to
separate the
% organize another
execution stream for the data packets reading from the main execution stream
for the X11 events handling. This choice has been
justified by the fact, that the BSD scheduling algorithm for processes instead
of threads is more mature, refined, and featured.
The \verb|-a| option was added
to the {\bfseries\itshape itGUI(1)} to
% visualize ADC histograms, which allows us
reduce visualization expenses by default, and visualize ADC histograms with
\verb|-a|. If the same
option is not supplied to {\bfseries\itshape itconf(1)}, the
{\bfseries\itshape inttarg(4)} module will be configured not to collect ADC
data for histogramming, because these data are currently not useful.
% useful first of all for the hardware testing.
Only the stepper motor temperature channel (ADC1) will be read once per cycle
(at EoB or final quantum). The corresponding value could be found at the end of
\verb|INTTARG_CYC_BEG| packet, which was enlarged by 2 bytes. This feature reduces
the CAMAC activity overhead per each time quantum and allows to watch the stepper
motor temperature also during the cycles with an inactive target.
All the mentioned software improvements were successfully tested during the
December 2011 Nuclotron run.

The full IntTarg CDAQ system data set in the packet stream form
could be provided online for clients on the remote hosts. A client could be
something like a
(sub)event builder ((Sub)EvB, see \cite{IsupJINRC10-34}) implemented in the
kernel context by, for example, the following {\itshape ngdp} graph:
\verb|ng_ksocket|$\to$\verb|ng_defrag|$\to$\verb|ng_em[s]|.
In the user context a client could be, for example, a pipe of the
{\bfseries\itshape hose(1)} utility (writer end) from the
{\itshape netpipes} package (see {\bfseries\itshape netpipes(1)}) and a some
read-only
version of the {\bfseries\itshape itGUI(1)} (reader end), which allows the user
to observe but not control the internal target.
Of course, users are free to implement their own clients using the BSD socket and
{\itshape ngdp} {\bfseries\itshape packet(3)} program interfaces (APIs).

Some minor updates of the IntTarg CDAQ are also possible in future.
%if the operating experience shows its necessity.
 A separate canvas for
the trajectory changes and calculations during the target walking could be added.
The magnetic field reading with the inactive target will be useful. The
target trajectory calculation algorithm has some annoying features, so it could
be revised more or less essentially. The 8th and 9th multiscaler input readings
could be prolonged after the final quantum.

\vspace*{-5mm}

\section{Conclusions}
\label{inttarg.concl}

\vspace*{-3mm}

\hspace*{4mm} The new control and data acquisition
system for the Nuclotron ITS
% internal target
has been implemented
% from scratch
using the {\itshape ngdp},
{\itshape camac}, and ROOT packages to allow easy network distribution and
integration.
%The CAMAC electronic
%% of the internal target station of the Nuclotron (LHEP, JINR)
%was revised essentially and the rare CAMAC modules with
%expired resource were eliminated.
%So, the goal of present work was successfully achieved:
The outdated both DOS software and underlying CAMAC and computer hardware
have been replaced.
The previously implemented {\bfseries\itshape targinfo(1)} server
%, which was ran under the SCAN DAQ as a workaround,
has been integrated into the IntTarg CDAQ now.
During 150 hours of the 2 Nuclotron runs the IntTarg CDAQ has demonstrated
the operation stability and target manipulation convenience.

\section*{Acknowledgements}
%\addcontentsline{toc}{section}{Acknowledgements}
\label{inttarg.Ack}

\vspace*{-3mm}

\hspace*{4mm} The author has
% authors have
a pleasure to thank S.G.Reznikov, who has designed and
implemented the hardware upgrade of the ITS
%internal target
control scheme on
the generic CAMAC
hardware, V.P.Ladygin --- for initiation of the presented developments,
\fbox{V.A.Krasnov} --- for useful discussions, S.M.Piyadin, A.N.Livanov and
A.N.Khrenov --- for support
during the run.
The present work was supported in part by the RFBR grant 10-02-00087a.

\vspace*{-5mm}

%\bibliography{preamble,online,sphere,inttarg}
%%camac_lnp
%\bibliographystyle{rusunsrt}
\newcommand{\noopsort}[1]{} \newcommand{\printfirst}[2]{#1}
  \newcommand{\singleletter}[1]{#1} \newcommand{\switchargs}[2]{#2#1}

\end{document}